\documentclass[a4paper,12pt]{article}
\usepackage{amssymb,amsmath}
\usepackage{amsfonts}
\usepackage{indentfirst}
\usepackage{graphicx}
\newcommand{\ds}{\displaystyle}
\title{Application of two spectral methods to a problem of convection with uniform internal heat source}
\author{Ioana Dragomirescu$^{*}$, Adelina Georgescu$^{**}$\\
{\small $^{*}$ Univ. "Politehnica" of Timisoara,\ Dept. of Mathematics }\\
{\small P-ta Victoriei, No.2, 300006, Timisoara, Romania}\\
 {\small $^{**}$ Univ. of Pitesti, Dept. of Mathematics }\\
{\small Str. Targu din Vale, No.1, 110040, Pitesti, Romania}}
\begin{document}
\date{}
\maketitle
\begin{abstract}
Two methods based on Fourier series expansions (a Chandrasekhar
functions-based method and a shifted Legendre polynomials -based
method) are used to study analytically the eigenvalue problem
governing the linear convection problem with an uniform internal
heat source in a horizontal fluid layer bounded by two rigid
walls. For each method some theoretical remarks are made.
Numerical results are given and they are compared with some
existing ones. Good agrement is found.
\end{abstract}
\par {\bf MSC: 76E06}
\par {\bf Keywords:} eigenvalue problem, convection, internal heat
source.
\section{Problem setting}
\par The effects of the presence in a fluid of an internal heat source have been experimentally,
numerically and analytically investigated by researchers in many convection problems \cite{Mi}, \cite{Ro},
\cite{Tr},\cite{Ve}. The investigations concerned the effects of the heating and cooling rate. Various conditions
were imposed on the lower and upper boundaries. The motion in the atmosphere or mantle convection are two among
phenomena of natural convection induced by internal heat sources. They bifurcate from the conduction state as a
result of its loss of stability. In spite of their importance, due to the occurrence of variable coefficients in
the nonlinear partial differential equations governing the evolution of the perturbations around the basic
equilibrium, so far these phenomena were treated mostly numerically and experimentally.
\par Herein a horizontal layer of viscous incompressible fluid with constant
viscosity and thermal conductivity coefficients $\nu$ and $k$ is considered \cite{Ve}.  In this context, the heat
and hydrostatic transfer equations  are \cite{Ve}
\begin{equation}
\label{eq:heat} \eta=k\dfrac{\partial^{2}\theta_{B}}{\partial
z^{2}},
\end{equation}
\begin{equation}
\label{eq:hydro} \dfrac{dp_{B}}{dz}=-\rho_{B}g,
\end{equation}
where $\eta=const.$ is the heating rate, $\theta_{B}$, $p_{B}$ and $\rho_{B}$ are the potential temperature,
pressure and density in the basic state, respectively. In the fluid, the temperature at all point varies at the
same rate as the boundary temperature, so the problem is characterized by a constant potential temperature
difference between the lower and the upper boundaries $\Delta\theta_{B}=\theta_{B_{0}}-\theta_{B_{1}}$. Taking
into account (\ref{eq:heat}) this leads to the following formula for the potential temperature distribution
\cite{Ve}
\begin{equation}
\label{eq:potential_t}
\theta_{B}=\theta_{B_{0}}-\dfrac{\Delta\theta_{B}}{h}\Big(z+\dfrac{h}{2}\Big)+\dfrac{\eta}{2k}\Big[z^{2}-
\Big(\dfrac{h^{2}}{2}\Big)^{2}\Big].
\end{equation}
\par  In nondimensional variables the system of equations characterizing the problem is
\begin{equation}
\label{eq:ecuatie_nond} \left\{
\begin{array}{l}
\dfrac{d{\bf U}}{dt}= -\nabla p'+\Delta {\bf U}+Gr \theta' {\bf k},\\
\\
\textrm{div} {\bf U}=0,\\
\\
\dfrac{d\theta'}{dt}=(1-Nz){\bf U}{\bf k}+Pr^{-1}\Delta \theta',
\end{array}
\right.
\end{equation}
where ${\bf U}=(u,v,w)$ is the velocity, $\theta'$ and $p'$ are the temperature
 and pressure deviations from the basic state \cite{Ve}, $Gr$ is the Grashof number, $Pr$ is the Prandtl number and
 $N$ is a nondimensional parameter characterizing the heating (cooling) rate of the layer.
 \par The boundaries are assumed rigid and ideal heat conducting, so the boundary conditions
 read
 \begin{equation}
 \label{eq:b_conditions}
 {\bf U}=\theta'=0 \textrm{ at } z=-\dfrac{1}{2} \textrm{ and } z=\dfrac{1}{2}.
 \end{equation}
 \par In \cite{Ve} the numerical investigations concerned the vertical distribution of the total heat fluxes and their
 individual components for small and moderate supercritical Rayleigh number in the presence of a uniform heat
 source.
 \par  The
 eigenvalue problem associated with the equations for a
convection problem with an uniform internal heat source in a horizontal fluid layer bounded by two rigid walls was
deduced in \cite{Dr2}.
\par
 Consider the viscous incompressible fluid confined into a periodicity rectangular box $V:0\leq x \leq a_{1}$,
  $ 0\leq y\leq a_{2}$, $-\dfrac{1}{2}\leq z\leq \dfrac{1}{2}$ \cite{Ge} bounded by two rigid horizontal walls. The corresponding eigenvalue problem \cite{Dr2}
  has the form
\begin{equation}
\label{eq:eigen1}
\left\{
\begin{array}{l}
(D^{2}-a^{2})^{2}W-a^{2}Ra \Theta=0,\\
(D^{2}-a^{2})\Theta+(1-Nz)W=0.
\end{array}
\right.
\end{equation}
with the boundary conditions
\begin{equation}
\label{eq:bc_eigen1} W=DW=\Theta=0 \textrm{ at } z=\pm \dfrac{1}{2}.
\end{equation}
\par In (\ref{eq:eigen1}) the Rayleigh number $Ra$ represents the eigenvalue while $(W, \Theta)$ represents
the corresponding eigenvector. The analytical study of this stability problem consists in finding the smallest
eigenvalue, i.e. the critical value of the Rayleigh number at which the convection sets in.
\par In \cite{Dr2} the analytical study of the eigenvalue problem (\ref{eq:eigen1})-(\ref{eq:bc_eigen1})
was performed by means of a method from \cite{Ch}. First the
system (\ref{eq:eigen1})-(\ref{eq:bc_eigen1}) was written in a
more convenient independent variable $x=z+\dfrac{1}{2}$. Then, two
methods (one based on Fourier series expansions of the unknown
functions and other a variational one) were used in order to find
the smallest eigenvalue. Here, the analytical study in also based
on Fourier series expansions of the unknown functions, but the
expansion functions satisfy all boundary conditions.
\par Taking into account the form of the boundary conditions two methods are used and, for each of them, some
analytical remarks on the chosen sets of expansion functions are presented.
\section{A method based on Chandrasekhar functions}
In this method, the unknown function $W$ is expanded upon a complete  set of orthogonal functions that satisfy all
boundary conditions $\Big(W=DW=0$ at $z=\pm \dfrac{1}{2}\Big)$ and then, from $(\ref{eq:eigen1})_{2}$ we find the
expression of the unknown function $\Theta$. Replacing these expansions in $(\ref{eq:eigen1})_{1}$ and imposing
the condition that the left-hand side of the obtained equation to be orthogonal to each function from the
expansion set, we obtain an algebraic system of equations which leads us to the secular equation, yielding the
critical value of the Rayleigh number.
 \par When the normal component of the velocity and its derivative
are zero at $z=-\dfrac{1}{2}$ and $z=\dfrac{1}{2}$, the classical
set of complete orthogonal functions that satisfy these conditions
are the Chandrasekhar sets of functions $\{C_{n}\}_{n\in
\mathbb{N}}$, $\{S_{n}\}_{n\in \mathbb{N}}$\cite{Ch}
\begin{equation}
\label{eq:Ch1} C_{n}(z)=\dfrac{\cosh \lambda_{n}z}{\cosh
\lambda_{n}/2}-\dfrac{\cos \lambda_{n}z}{\cos \lambda_{n}/2},
\end{equation}
\begin{equation}
\label{eq:Ch2}
S_{n}(z)=\dfrac{\sinh(\mu_{n}z)}{\sinh(\mu_{n}/2)}-\dfrac{\sin
(\mu_{n}z)}{\sin(\mu_{n}/2)}
\end{equation} where  $\lambda_{n}$ and $\mu_{n}$ are the positive
roots of the equations $\tanh\Big(\dfrac{\lambda}{2}\Big)+\tan \Big(\dfrac{\lambda}{2}\Big)=0$ and
$\coth\Big(\dfrac{\mu}{2}\Big)-\cot\Big(\dfrac{\mu}{2}\Big)=0$. We have
$$\ds\int_{-0.5}^{0.5}C_{n}(z)C_{m}(z)dz=\ds\int_{-0.5}^{0.5}S_{n}(z)S_{m}(z)dz=\delta_{mn}.$$
\par By definition, the functions $C_{n}$ and $S_{n}$  and their derivatives vanish at $x=\pm \dfrac{1}{2}$
so the boundary conditions (\ref{eq:bc_eigen1}) are satisfied.
\par Let us consider
$W=\sum\limits_{n=1}^{\infty}W_{n}C_{n}(z)$. From $(\ref{eq:eigen1})_{2}$ we obtain the expression of the unknown
function $\Theta$,
$$
\begin{array}{l}
\Theta=A\cosh az+B\sinh az+\dfrac{W_{n}\cosh
\lambda_{n}z(Nz-1)}{(\lambda_{n}^{2}-a^{2})\cosh
\lambda_{n}/2}-\dfrac{2\lambda_{n}NW_{n}}{(\lambda_{n}^{2}-a^{2})^{2}\cosh
\lambda_{n}/2}\cdot \\
\cdot\sinh \lambda_{n}z+\dfrac{(1-Nz)W_{n}\cos \lambda_{n}z}{(\lambda_{n}^{2}+a^{2})\cos
\lambda_{n/2}}-\dfrac{2\lambda_{n}NW_{n}}{(\lambda_{n}^{2}+a^{2})^{2}\cos \lambda_{n}/2}\sin \lambda_{n}z,
\end{array}
$$
where
$A=\dfrac{2a^{2}W_{n}}{(\lambda_{n}^{2}-a^{2})(\lambda_{n}^{2}+a^{2})\cosh
a/2}$  and
$$B=\dfrac{8\lambda_{n}^{3}NW_{n}a^{2}}{(\lambda_{n}^{2}-a^{2})^{2}(\lambda_{n}^{2}+a^{2})^{2}\cosh
\lambda_{n}/2}-\dfrac{a^{2}NW_{n}}{(\lambda_{n}^{2}-a^{2})(\lambda_{n}^{2}+a^{2})}.$$
\par
However, in our case, replacing these expressions in $(\ref{eq:eigen1})_{1}$ and imposing the condition that the
left-hand side of the obtained equation to be orthogonal to $C_{m}$, $m\in \mathbb{N}$, we obtain an expression in
which the physical parameter $N$ is missing. The mathematical explanation is that the chosen set of expansion
functions introduced an extraparity (inexistent in the given problem), leading to the loss of one of the physical
parameter, in this case the cooling (heating) rate $N$.
\par {\bf Remark.}   The physical parameter $N$ also disappear when the expansion functions are
$S_{n}$, $n=1,2,...$.
\par Another explanation could be the fact that we have no physical or mathematical reason to assume that $W$ is
either even or odd. The general form of $W$ will be considered elsewhere.
\section{A method based on shifted Legendre polynomials}
\par In order to avoid the loss of $N$, we use a different set of
orthogonal functions, namely a basis of shifted Legendre polynomials (SLP) on $[0,1]$.
\par Let us modify the system (\ref{eq:eigen1}) by a translation of the variable $z$,
$x=z+\dfrac{1}{2}$, such that the eigenvalue problem becomes
\begin{equation}
\label{eq:eigen2} \left\{
\begin{array}{l}
(D^{2}-a^{2})^{2}W-a^{2}Ra \Theta=0,\\
(D^{2}-a^{2})\Theta+(N_{1}-Nx)W=0,
\end{array}
\right.
\end{equation}
with $N_{1}=1+\dfrac{N}{2}$ and the boundary conditions
\begin{equation}
\label{eq:eigen_bc3} W=DW=\Theta=0 \textrm{ at }x=0 \textrm{ and
}1.
\end{equation}
\par Starting with the classical Legendre polynomials defined on $(-1,1)$,
let us introduce the complete sets of expansion functions. We are
interes-ted in expansion functions that satisfy all boundary
conditions.  Let
 $H^{1}_{0}(0,1)$, $H^{2}_{0}(0,1)$ be two Hilbert spaces \cite{HSTR}
 $$H_{0}^{1}(0,1)=\{f| f, f'\in L^{2}(0,1), f(0)=f(1)=0 \}, $$
 $$H^{2}_{0}(0,1)=\{f|f,f',f''\in L^{2}(0,1), f(0)=f(1)=f'(0)=f'(1)=0\}$$
and denote by $L_{k}$ the Legendre polynomials defined on
$(-1,1)$. By means of them, we construct the SLP (denoted by us by
$Q_{k}$) on $(a,b)$, namely
$Q_{k}(x)=L_{k}\Big(\dfrac{2x-a-b}{b-a}\Big).$ Taking
$(a,b)=(0,1)$ we find that $Q_{k}$ are orthogonal polynomials on
the interval $(0,1)$, i.e. $
\ds\int_{0}^{1}Q_{i}Q_{j}dz=\dfrac{1}{2i+1}\delta_{ij}$. Using the
identity \cite{HSTR}

\begin{equation}
\label{eq:rec_der} 2(2i+1)Q_{i}(z)=Q'_{i+1}(z)-Q'_{i-1}(z).
\end{equation}
we define the complete sets of orthogonal functions $\{\phi_{i}\}_{i=1,2,...}\subset H_{0}^{1}(0,1)$,
$$\phi_{i}(z)=\ds\int_{0}^{z}Q_{i}(t)dt=\dfrac{Q_{i+1}-Q_{i-1}}{2(2i+1)},$$
satisfying boundary conditions $\phi_{i}(0)=\phi_{i}(1)=0$ at $z=0$ and $1$ and $\{\beta_{i}\}_{i=1,2,...}\subset
H_{0}^{2}(0,1)$,
$$\beta_{i}(z)=\ds\int_{0}^{z}\ds\int_{0}^{s}Q_{i+1}(t)dt
ds=\dfrac{1}{4}\Big[\dfrac{Q_{i+3}-Q_{i+1}}{(2i+3)(2i+5)}-\dfrac{Q_{i+1}-Q_{i-1}}{(2i+1)(2i+3)}\Big],$$ satisfying
boundary conditions $\beta_{i}(0)=\beta_{i}(1)=\beta'_{i}(0)=\beta'_{i}(1)=0$ at $z=0$ and $1$.
\par {\bf Remark.} We could also work with SLP on $(a,b)=\Big(-\dfrac{1}{2}, \dfrac{1}{2}\Big)$. However, the choice $(a,b)=(0,1)$ leads us to
simplified numerical evaluations.
\par The system (\ref{eq:eigen1}) can be
solved numerically by approximating the solution $(W,\Theta)$ by
\begin{equation}
\label{eq:dezv} W=\sum\limits_{i=1}^{n} W_{i}\beta_{i}(z), \  \
\Theta=\sum\limits_{i=1}^{n}\Theta_{i}\phi_{i}(z)
\end{equation}
with $W_{i}$ and $\Theta_{i}$ the Fourier coefficients. In this
way, the system (\ref{eq:eigen1}) can be written in terms of the
expansion functions only
\begin{equation}
\label{eq:sistem} \left\{
\begin{array}{l}
\sum\limits_{i=1}^{n}[W_{i}(D^{2}-a^{2})^{2}\beta_{i}-a^{2}Ra \Theta_{i}\phi_{i}]=0,\\
\sum\limits_{i=1}^{n}[\Theta_{i}(D^{2}-a^{2})\phi_{i}+(N_{1}-Nz)W_{i}\beta_{i}]=0.
\end{array}
\right.
\end{equation}
Multiplying the system (\ref{eq:sistem}) by the vector $(\beta_{k}, \phi_{k})$ we obtain the algebraic system
\begin{equation}
\label{eq:sistem_orto} \left\{
\begin{array}{l}
\sum\limits_{i=1}^{n}[W_{i}\Big((D^{2}-a^{2})^{2}\beta_{i}, \beta_{k}\Big)-a^{2}Ra \Theta_{i}(\phi_{i}, \beta_{k})]=0,\\
\sum\limits_{i=1}^{n}[\Theta_{i}\Big((D^{2}-a^{2})\phi_{i}, \phi_{k}\Big)+W_{i}N_{1}(\beta_{i},
\phi_{k})-W_{i}N(z\beta_{i},\phi_{k})]=0.
\end{array}
\right.
\end{equation}
Taking into account the fact that the coefficients $W_{i}$, $\Theta_{i}$ are not all null, i.e. the Cramer
determinant vanishes, the secular equation has the form
\begin{equation}
\label{eq: ec_sec}
\begin{tabular}{|cc|}
$((D^{2}-a^{2})^{2}\beta_{i},\beta_{k})$&$-a^{2}Ra (\phi_{i},\beta_{k})$\\
$N_{1}(\beta_{i},\phi_{k})-N(z\beta_{i}, \phi_{k})$&$((D^{2}-a^{2})\phi_{i}, \phi_{k})$
\end{tabular}=0.
\end{equation}
 The scalar products from (\ref{eq: ec_sec}) are given in the Appendix. \par The system (\ref{eq:eigen2}) has variable coefficients (functions of $x$). In this case, the following
recurrence relation was used for the numerical study
\begin{equation}
\label{eq:rec_z} 2zQ_{i}=\dfrac{i+1}{2i+1}Q_{i+1}+Q_{i}+\dfrac{i}{2i+1}Q_{i-1}.
\end{equation}
\section{Numerical results}
Taking $n=m=1$ we obtained a first approximation of the Rayleigh number, which proved to be a good approximation
compared to the one obtained in $\cite{Dr2}$. The obtained numerical results are presented in Table 1 in
comparison with the results from \cite{Dr2}. The disadvantage of this method is given by the fact that the
approximations are limited by the difficult evaluation of the associated matrix for a large number of functions in
the expansion sets.  However, the expressions of the neutral manifolds are easy to obtain with this method. For a
large number of terms in the Fourier series expansions, we must use an algorithm for solving the algebraic
equation (\ref{eq: ec_sec}). For instance in \cite{HSTR}, the Arnoldi algorithm is used.

\bigskip

\begin{center}
\begin{tabular}{|c|c|c|c|c|}
\hline $N$&$a^{2}$&$Ra- Fourier$&$R_{a} - var. meth.$&$Ra- Legendre$\\
\hline $0$&$9.711$&$1715.079324$&$1749.97575$&$1749.95727$\\
\hline $1$&$9.711$&$1711.742588$&$1746.804944$&$1746.809422$\\
\hline $2$&$9.711$&$1701.891001$&$1737.45025$&$1737.450242$\\
\hline $1$&$10.0$&$1712.257687$&$1747.29100$& $1747.290998$\\
\hline $4$ &$10.0$&$1664.341789$&$1701.62704$&$1701.627037$\\
\hline $4$&$12.0$&$1685.422373$&$1723.62407$&$1723.624047$\\
\hline $8$&$12.0$&$1547.460446$&$1590.19681$&$1590.196769$\\
\hline $9$&$12.0$&$1508.147637$&$1551.72378$&$1551.723746$\\
\hline $10$&$12.0$&$1468.449223$&$1512.69203$&$1512.691998$\\
\hline $12$ &$12$&$1389.837162$&$1434.90396$&$1434.903926$\\
\hline $16$&$12$&$1243.442054$&$1288.50149$&$1288.501459$\\
\hline $10$&$9.0$&$1482.527042$&$1525.59302$&$1525.593072$\\
\hline $11$&$9.0$&$1446.915467$&$1490.55802$&$1490.558078$\\
\hline $12$&$9.00$&$1411.401914$&$1455.48233$&$1455.482384$\\
\hline
\end{tabular}\end{center}
\begin{center}
{\bf Table 1. } Numerical evaluations of the Rayleigh number for various values of the parameters $N$ and $a$.
\end{center}
\par When the wavenumber is kept constant an increase in the heating (cooling) rate parameter leads to a decreasing
of the Rayleigh number. When $N=0$ the problem reduces to the particular case of Rayleigh-B\'{e}nard convection
and the numerical evaluation lead us to a value similar to the classical value for the Rayleigh number, i.e.
$Ra=1749.95727$ for $a=3.117$.

\section{Appendix}
Let us give the expressions of the scalar products occurring in (\ref{eq: ec_sec}). Since in
$(\ref{eq:eigen2})_{1}$ the expression $((D^{2}-a^{2})^{2}\beta_{i}, \beta_{k})$ is written as
$$((D^{2}-a^{2})^{2}\beta_{i}, \beta_{k})=(D^{4}\beta_{i}, \beta_{k})-2a^{2}(D^{2}\beta_{i}, \beta_{k})+a^{4}(\beta_{i},
\beta_{k})$$ let us simplify these products or simply evaluate them. Taking into account the definition of the
scalar product on $L^{2}(0,1)$, i.e. $(f,g)=\ds\int_{0}^{1}fg dz$ and the boundary conditions satisfied by the
expansion functions, we have
\begin{equation}
\label{eq:ik1} {\small (D^{4}\beta_{i}, \beta_{k})=(\beta''_{i}, \beta''_{k})=\left\{
\begin{array}{l}
\dfrac{1}{2i+3}  \textrm{ if } i=k,\\
\\
0 \textrm{ if } i\neq k
\end{array}
\right.}
\end{equation}
and
\begin{equation}
\label{eq:ik2} {\small (D^{2}\beta_{i},\beta_{k})=-(\beta'_{i}, \beta'_{k})=\left\{
\begin{array}{l}
-\dfrac{1}{2(2i+1)(2i+3)(2i+5)} \textrm{ if } i=k,\\
\\
\dfrac{1}{4(2i-1)(2i+1)(2i+3)} \textrm{ if } i=k+2,\\
\\
0 \textrm { otherwise }
\end{array}
\right.}
\end{equation}
Given the fact that
$(\beta_{i},\beta_{k})=\dfrac{1}{2}\Big(\dfrac{\phi_{i+2}-\phi_{i}}{2i+3},\dfrac{\phi_{k+2}-\phi_{k}}{2k+3}\Big)$
we first evaluated the product $(\phi_{i}, \phi_{k})$ and we get
\begin{equation}
\label{eq:ik6} {\small (\phi_{i},\phi_{k})=\left\{
\begin{array}{l}
\dfrac{1}{2(2i-1)(2i+1)(2i+3)} \textrm{ if } i=k,\\
\\
-\dfrac{1}{4(2i+1)(2i+3)(2i+5)}  \textrm{ if } i=k-2,\\
\\
0 \textrm{ otherwise }
\end{array}
\right.}
\end{equation} Using (\ref{eq:ik6}) we have
\begin{equation}
\label{eq:ik3} (\beta_{i},\beta_{k})= {\small
\begin{array}{l}
\\
\left\{
\begin{array}{l}
\dfrac{3}{8(2i-1)(2i+1)(2i+3)(2i+5)(2i+7)} \textrm{ if } i=k,\\
\\
-\dfrac{1}{4(2i+1)(2i+3)(2i+5)(2i+7)(2i+9)}  \textrm{ if } i=k-2,\\
\\
\dfrac{1}{16(2i+3)(2i+5)(2i+7)(2i+9)(2i+11)} \textrm{ if } i=k-4, \\
\\
0 \textrm{ otherwise }
\end{array}
\right.
\end{array}}
\end{equation}
We also used (\ref{eq:ik6}) to deduce $(\phi_{i}, \beta_{k})$, i.e.
\begin{equation}
\label{eq:ik4} {\small \begin{array}{l}
(\phi_{i},\beta_{k})=\dfrac{1}{2(2k+3)}[(\phi_{i},\phi_{k+2})-(\phi_{i},\phi_{k})]=\\
=\left\{
\begin{array}{l}-\dfrac{3}{8(2i-1)(2i+1)(2i+3)(2i+5)} \textrm{ if } i=k,\\
\\
\dfrac{3}{8(2i-3)(2i-1)(2i+1)(2i+3)} \textrm{ if }i=k+2,\\
\\
\dfrac{1}{8(2i+1)(2i+3)(2i+5)(2i+7)}  \textrm{ if } i=k-2,\\
\\
-\dfrac{1}{8(2i-5)(2i-3)(2i-1)(2i+1)} \textrm{ if } i=k+4, \\
\\
0 \textrm{ otherwise }
\end{array}
\right.
\end{array}}
\end{equation}
Let us remark that $(\beta_{i},\phi_{k})=(\phi_{k},\beta_{i})$.
\par The computations of $(D^{2}\phi_{i}, \phi_{k})$ was simplified by the expressions of the $\phi_{i}$ functions.
We have
\begin{equation}
\label{eq:ik5}{\small (D^{2}\phi_{i}, \phi_{k})=-(Q_{i}, Q_{k})=\left\{
\begin{array}{l}
-\dfrac{1}{2i+1} \textrm{ if }i=k,\\
\\
0 \textrm { otherwise }
\end{array}.
\right.}
\end{equation}
All the obtained expressions (\ref{eq:ik1}) - (\ref{eq:ik8}) are based on the orthogonality relationship between
the SLP. In deducing the expression below we also used the recurrence relation (\ref{eq:rec_z})
\begin{equation}
\label{eq:ik8} {\small (z\beta_{i},\phi_{k})=\left\{
\begin{array}{l}
-\dfrac{i+4}{16(2i+3)(2i+5)(2i+7)(2i+9)(2i+11)} \textrm{ if } i=k-5,\\
\\
-\dfrac{1}{16(2i+3)(2i+5)(2i+7)(2i+9)} \textrm{ if }i=k-4,\\
\\
\dfrac{1}{16(2i+1)(2i+3)(2i+5)(2i+9)}  \textrm{ if } i=k-3,\\
\\
\dfrac{3}{16(2i+1)(2i+3)(2i+5)(2i+7)} \textrm{ if } i=k-2,\\
\\
-\dfrac{3}{16(2i-1)(2i+1)(2i+3)(2i+5)(2i+7)} \textrm{ if } i=k-1,\\
\\
-\dfrac{3}{16(2i-1)(2i+1)(2i+3)(2i+5)} \textrm{ if } i=k,\\
\\
-\dfrac{1}{16(2i-3)(2i+1)(2i+3)(2i+5)} \textrm{ if } i=k+1,\\
\\
\dfrac{1}{16(2i-3)(2i-1)(2i+1)(2i+3)} \textrm{ if} i=k+2,\\
\\
\dfrac{i+1}{16(2i-5)(2i-3)(2i-1)(2i+1)(2i+3)} \textrm { if }
i=k+3\\
\\
0 \textrm{ otherwise }
\end{array}
\right.}
\end{equation}
\section{Conclusions}
In this paper we performed an analytical study of the eigenvalue
problem corresponding to a convection problem with uniform
internal heat source. We pointed out some aspects of the spectral
methods that we employed concerning the sets of the expansion
functions that can be used to an analytical study of this problem.
As in this case the expansion sets of Chandrasekhar functions
introduced an extraparity they were not appropriate. However, for
some other problems \cite{Dr3} their use proved to be successful.
The method based on SLP lead to good numerical approximations. All
numerical results obtained with this method are compared with the
existing ones. The effect of the heating (cooling) rate on the
values of the Rayleigh number is pointed out.

\end{document}